\documentclass[DIV12,11pt]{scrartcl}

\usepackage{amssymb,amsmath,amsthm}
\usepackage{cite,graphicx}
\usepackage{authblk}

\setkomafont{title}{\normalfont \LARGE}
\setkomafont{section}{\normalfont \Large \bfseries \boldmath}
\setkomafont{subsection}{\normalfont \large \bfseries\boldmath}
\setkomafont{subsubsection}{\normalfont \normalsize \bfseries}
\setkomafont{descriptionlabel}{\itshape}
\setkomafont{paragraph}{\normalfont \bfseries \boldmath}

\DeclareMathOperator{\probab}{\mathbb{P}}
\DeclareMathOperator{\expected}{\mathbb{E}}

\newcommand{\real}{\mathbb{R}}
\newcommand{\eps}{\varepsilon}

\newtheorem{theorem}{Theorem}[section]
\newtheorem{lemma}[theorem]{Lemma}

\newtheorem{corollary}[theorem]{Corollary}

\title{Smoothed Analysis of the Minimum-Mean Cycle Canceling Algorithm and the Network Simplex Algorithm}

\author[1]{Kamiel Cornelissen}
\author[1]{Bodo Manthey}

\affil[1]{University of Twente, Department of Applied Mathematics \authorcr
\texttt{k.cornelissen@utwente.nl, b.manthey@utwente.nl}}

\date{\vspace{-5ex}}
\begin{document}

\maketitle

\begin{abstract}
The minimum-cost flow (MCF) problem is a fundamental optimization problem with many applications and seems to be well understood.
Over the last half century many algorithms have been developed to solve the MCF problem and these algorithms have varying worst-case bounds on their running time. However, these worst-case bounds are not always a good indication of the algorithms' performance in practice. The Network Simplex (NS) algorithm needs an exponential number of iterations for some instances, but it is considered the best algorithm in practice and performs best in experimental studies. On the other hand, the Minimum-Mean Cycle Canceling (MMCC) algorithm is strongly polynomial, but performs badly in experimental studies. 

To explain these differences in performance in practice we apply the framework of smoothed analysis. We show an upper bound of $O(mn^2\log(n)\log(\phi))$ for the number of iterations of the MMCC algorithm. Here $n$ is the number of nodes, $m$ is the number of edges, and $\phi$ is a parameter limiting the degree to which the edge costs are perturbed. We also show a lower bound of $\Omega(m\log(\phi))$ for the number of iterations of the MMCC algorithm, which can be strengthened to $\Omega(mn)$ when $\phi=\Theta(n^2)$. For the number of iterations of the NS algorithm we show a smoothed lower bound of $\Omega(m \cdot \min \{ n, \phi \} \cdot \phi)$.
\end{abstract}

\section{Introduction}
The minimum-cost flow (MCF) problem is a well-studied problem with many applications, for example, modeling transportation and communication networks~\cite{AhujaNetworkFlow,ForFul62}. Over the last half century many algorithms have been developed to solve it. The first algorithms proposed in the 1960s were all pseudo-polynomial. These include the Out-of-Kilter algorithm by Minty~\cite{Min60} and by Fulkerson~\cite{Ful61}, the Cycle Canceling algorithm by Klein~\cite{Kle67}, the Network Simplex (NS) algorithm by Dantzig~\cite{DantzigNS}, and the Successive Shortest Path (SSP) algorithm by Jewell~\cite{Jew62}, Iri~\cite{Iri60}, and Busacker and Gowen~\cite{BusGow60}. In 1972 Edmonds and Karp~\cite{DBLP:journals/jacm/EdmondsK72} proposed the Capacity Scaling algorithm, which was the first polynomial MCF algorithm. In the 1980s the first strongly polynomial algorithms were developed by Tardos~\cite{DBLP:journals/combinatorica/Tardos85} and by Orlin~\cite{Orl84}. Later, several more strongly polynomial algorithms were proposed such as the Minimum-Mean Cycle Canceling (MMCC) algorithm by Goldberg and Tarjan~\cite{GoldbergTarjan} and the Enhanced Capacity Scaling algorithm by Orlin~\cite{Orl93}, which currently has the best worst-case running time. For a more complete overview of the history of MCF algorithms we refer to Ahuja et al.~\cite{AhujaNetworkFlow}.  

When we compare the performance of several MCF algorithms in theory and in practice, we see that the algorithms that have good worst-case bounds on their running time are not always the ones that perform best in practice. Zadeh~\cite{Zadeh} showed that there exist instances for which the Network Simplex (NS) algorithm has exponential running time, while the Minimum-Mean Cycle Canceling (MMCC) algorithm runs in strongly polynomial time, as shown by Goldberg and Tarjan~\cite{GoldbergTarjan}. In practice however, the relative performance of these algorithms is completely different. Kov\'acs~\cite{KovacsExperimental}
showed in an experimental study that the NS algorithm is much faster than the MMCC algorithm on practical instances. In fact, the NS algorithm is even the fastest MCF algorithm of all. An explanation for the fact that the NS algorithm performs much better in practice than indicated by its worst-case running time is that the instances for which it needs exponential time are very contrived and unlikely to occur in practice. To better understand the differences between worst-case and practical performance for the NS algorithm and the MMCC algorithm, we analyze these algorithms in the framework of smoothed analysis.   

Smoothed analysis was introduced by Spielman and Teng~\cite{SpielmanTeng} to explain why the simplex algorithm usually needs only a polynomial number of iterations in practice, while in the worst case it needs an exponential number of iterations. In the framework of smoothed analysis, an adversary can specify any instance and this instance is then slightly perturbed before it is used as input for the algorithm. This perturbation can model, for example, measurement errors or numerical imprecision. In addition, it can model noise on the input that can not be quantified exactly, but for which there is no reason to assume that it is adversarial. Algorithms that have a good smoothed running time often perform well in practice. We refer to two surveys~\cite{it,SpielmanT09} for a summary of results that have been obtained using smoothed analysis.  
 
We consider a slightly more general model of smoothed analysis, introduced by Beier and V{\"o}cking~\cite{BV04}. In this model the adversary can not only specify the mean of the noisy parameter, but also the type of noise. We use the following smoothed input model for the MCF problem. An adversary can specify the structure of the flow network including all nodes and edges, and also the exact edge capacities and budgets of the nodes. However, the adversary can not specify the edge costs exactly. For each edge $e$ the adversary can specify a probability density $g_e: [0,1]\rightarrow [0,\phi]$ according to which the cost of $e$ is drawn at random. The parameter $\phi$ determines the maximum density of the density function and can therefore be interpreted as the power of the adversary. If $\phi$ is large, the adversary can very accurately specify each edge cost and we approach worst-case analysis. If $\phi=1$, the adversary has no choice but to specify the uniform density on the interval $[0,1]$ and we have average-case analysis. 

Brunsch et al.~\cite{BrunschSSP} were the first to show smoothed bounds on the running time of an MCF algorithm. They showed that the SSP algorithm needs $O(mn\phi)$ iterations in expectation and has smoothed running time $O(mn\phi(m+n\log(\phi)))$, since each iteration consists of finding a shortest path. They also provide a lower bound of $\Omega(m\cdot\min\{n,\phi\}\cdot\phi)$ for the number of iterations that the SSP algorithm needs, which is tight for $\phi= \Omega(n)$. These bounds show that the SSP algorithm needs only a polynomial number of iterations in the smoothed setting, in contrast to the exponential number it needs in the worst case, and explains why the SSP algorithm performs quite well in practice. In order to fairly compare the SSP algorithm with other MCF algorithms in the smoothed setting, we need smoothed bounds on the running times of these other algorithms.
Brunsch et al.~\cite{BrunschSSP} asked particularly for smoothed running time bounds for the MMCC algorithm,
since the MMCC algorithm has a much better worst-case running time than the SSP algorithm, but performs worse in practice.
It is also interesting to have smoothed bounds for the NS algorithm, since the NS algorithm is the fastest MCF algorithm in practice. However, until now no smoothed bounds were known for other MCF algorithms. In this paper we provide smoothed lower and upper bounds for the MMCC algorithm, and a smoothed lower bound for the NS algorithm.

For the MMCC algorithm we provide an upper bound (Section~\ref{UBMMCC}) for the expected number of iterations that the MMCC algorithm needs of $O(mn^2\log(n)\log(\phi))$. For dense graphs, this is an improvement over the $\Theta(m^2n)$ iterations that the MMCC algorithm needs in the worst case, if we consider $\phi$ a constant (which is reasonable if it models, for example, numerical imprecision or measurement errors). 

We also provide a lower bound (Section~\ref{LBConstantPhi}) on the number of iterations that the MMCC algorithm needs. For every $n$, every $m\in\{n,n+1,\ldots,n^2\}$, and every $\phi\leq2^n$, we provide an instance with $\Theta(n)$ nodes and $\Theta(m)$ edges for which the MMCC algorithm requires $\Omega(m\log(\phi))$ iterations. For $\phi=\Omega(n^2)$ we can improve our lower bound (Section~\ref{LBBigPhi}). We show that for every $n\geq 4$ and every $m\in\{n,n+1,\ldots,n^2\}$, there exists an instance with $\Theta(n)$ nodes and $\Theta(m)$ edges, and $\phi=\Theta(n^2)$, for which the MMCC algorithm requires $\Omega(mn)$ iterations. This is indeed a stronger lower bound than the bound for general $\phi$, since we have $m\log(\phi)=\Theta(m\log(n))$ for $\phi=\Theta(n^2)$.    

For the NS algorithm we provide a lower bound (Section~\ref{LBNS}) on the number of non-degenerate iterations that it requires. In particular, we show that for every $n$, every $m \in \{ n, \ldots, n^2 \}$, and every $\phi \leq 2^n$ there exists a flow network with $\Theta(n)$ nodes and $\Theta(m)$ edges, and an initial spanning tree structure for which the NS algorithm needs $\Omega(m\cdot \min\{n,\phi\}\cdot \phi)$ non-degenerate iterations with probability $1$. The existence of an upper bound is our main open problem.
Note that our bound is the same as the lower bound that Brunsch et al.~\cite{BrunschSSP} found for the smoothed number of iterations of the SSP algorithm. This is no coincidence, since we use essentially the same instance (with some minor changes) to show our lower bound. We show that with the proper choice of the initial spanning tree structure for the NS algorithm, we can ensure that the NS algorithm performs the same flow augmentations as the SSP algorithm and therefore needs the same number of iterations (plus some degenerate ones).   

In the rest of our introduction we introduce the MCF problem, the MMCC algorithm and the NS algorithm in more detail. In the rest of our paper, all logarithms are base $2$.

\subsection{Minimum-Cost Flow Problem} \label{IntroMCF}
A \emph{flow network} is a simple directed graph $G=(V,E)$ together with a nonnegative capacity function $u:E\rightarrow \real_{+}$ defined on the edges. For convenience, we assume that $G$ is connected and that $E$ does not contain a pair $(u,v)$ and $(v,u)$ of reverse edges. For the MCF problem, we also have a cost function $c:E\rightarrow [0,1]$ on the edges and a budget function $b:V\rightarrow \real$ on the nodes. Nodes with negative budget require a resource, while nodes with positive budget offer it. A \emph{flow} $f:E\rightarrow\real_{+}$ is a nonnegative function on the edges that satisfies the capacity constraints, $0\leq f(e) \leq u(e)$ (for all $e \in E$), and flow conservation constraints
$
  b(v) + \sum_{e = (u, v) \in E} f(e) = \sum_{e' = (v, w) \in E} f(e')$ (for all $v\in V$). The cost $c(f)$ of a flow $f$ is defined as the sum of the flow on each edge times the cost of that edge, that is, $c(f)=\sum_{e\in E} c(e)\cdot f(e)$. The objective of the minimum-cost flow problem is to find a flow of minimum cost or conclude that no feasible flow exists.
	
In our analysis we often use the concept of a \emph{residual network}, which we define here. For an edge $e=(u,v)$ we denote the reverse edge $(v,u)$ by $e^{-1}$. For flow network $G$ and flow $f$, the residual network $G_f$ is defined as the graph $G_f=(V, E_f\cup E_b)$. Here $E_f = \bigl\{ e \mid e \in E \text{ and } f(e) < u(e) \bigr\}$ is the set of \emph{forward edges} with capacity $u'(e)= u(e)-f(e)$ and cost $c'(e)=c(e)$. $E_b = \bigl\{ e \mid e^{-1} \in E \text{ and } f(e^{-1}) > 0 \bigr\}$ is the set of \emph{backward edges} with capacity $u'(e) = f(e^{-1})$ and cost $c'(e)=-c(e^{-1})$. Here $u'(e)$ is also called the \emph{residual capacity} of edge $e$ for flow $f$.  

\subsection{Minimum-Mean Cycle Canceling Algorithm}
The MMCC algorithm works as follows:
\begin{itemize}
\item First we find a feasible flow using any maximum-flow algorithm.
\item Next, as long as the residual network contains cycles of negative total cost, we find a cycle of minimum-mean cost and maximally augment flow along this cycle.
\item We stop when the residual network does not contain any cycles of negative total cost.
\end{itemize}

For a more elaborate description of the MMCC algorithm, we refer to Korte and Vygen~\cite{KorteVygen}. In the following, we denote the mean cost of a cycle $C$ by $\mu(C)=\left(\sum_{e\in C}c(e)\right)/|C|$. Also, for any flow $f$, we denote the mean cost of the cycle of minimum-mean cost in the residual network $G_f$ by $\mu(f)$.

Goldberg and Tarjan~\cite{GoldbergTarjan} proved in 1989 that the Minimum-Mean-Cycle Canceling algorithm runs in strongly polynomial time. Five years later Radzik and Goldberg~\cite{RadzikGoldberg} slightly improved this bound on the running time and showed that it is tight. In the following we will focus on the number of iterations the MMCC algorithm needs, that is, the number of cycles that have to be canceled. A bound on the number of iterations can easily be extended to a bound on the running time, by noting that a minimum-mean cycle can be found in $O(nm)$ time, as shown by Karp~\cite{KarpMMC}. The tight bound on the number of iterations that the MMCC algorithm needs is as follows.

\begin{theorem}[Radzik and Goldberg] \label{StrPonBound} 
The number of iterations needed by the MMCC algorithm is bounded by $O(nm^2)$ and this bound is tight.
\end{theorem}

To prove our smoothed bounds in the next sections, we use another result by Korte and Vygen~\cite[Corollary~9.9]{KorteVygen} which states that the absolute value of the mean cost of the cycle that is canceled by the MMCC algorithm, $|\mu(f)|$, decreases by at least a factor $1/2$ every $mn$ iterations. 

\begin{theorem}[Korte and Vygen] \label{LogBound}
Every $mn$ iterations of the MMCC algorithm, $|\mu(f)|$ decreases by at least a factor $1/2$.
\end{theorem}

%

\subsection{Network Simplex Algorithm}
The Network Simplex (NS) algorithm starts with an initial spanning tree structure $(T,L,U)$ and associated flow $f$, where each edge in $E$ is assigned to exactly one of $T$, $L$, and $U$, and it holds that
\begin{itemize}
\item $f(e)=0$ for all edges $e\in L$,
\item $f(e)=u(e)$ for all edges $e\in U$,
\item $0\leq f(e) \leq u(e)$ for all edges $e\in T$,
\item the edges of $T$ form a spanning tree of $G$ (if we consider the undirected version of both the edges of $T$ and the graph $G$).
\end{itemize}
If the MCF problem has a feasible solution, such a structure can always be found by first finding any feasible flow and then augmenting flow along cycles consisting of only edges that have a positive amount of flow less than their capacity, until no such cycles remain. Note that the structure $(T,L,U)$ uniquely determines the flow $f$, since the edges in $T$ form a tree. In addition to the spanning tree structure, the NS algorithm also keeps track of a set of node potentials $\pi(v)$ for all nodes $v\in V$. The node potentials are defined such that the potential of a specified root node is $0$ and that the potential for other nodes is such that the reduced cost $c^{\pi}(u,v)=c(u,v)-\pi(u)+\pi(v)$ of an edge $(u,v)$ equals $0$ for all edges $(u,v)\in T$.

In each iteration, the NS algorithm tries to improve the current flow by adding an edge to $T$ that violates its optimality condition. An edge in $L$ violates its optimality condition if it has strictly negative reduced cost, while an edge in $U$ violates its optimality condition if it has strictly positive reduced cost. One of the edges $e$ that violates its optimality condition is added to $T$, which creates a unique cycle $C$ in $T$. Flow is maximally augmented along $C$, until the flow on one of the edges $e'\in C$ becomes $0$ or reaches its capacity. The edge $e'$ leaves $T$, after which $T$ is again a spanning tree of $G$. Next we update the sets $T$, $L$, and $U$, the flow and the node potentials. This completes the iteration. If any edges violating their optimality condition remain, another iteration is performed. One iteration of the NS algorithm is also called a pivot. The edge $e$ that is added to $T$ is called the entering edge and the edge $e'$ that leaves $T$ is called the leaving edge. Note that in some cases the entering edge can be the same edge as the leaving edge. Also, if one of the edges in the cycle $C$ already contains flow equal to its capacity, the flow is not changed in that iteration, but the spanning tree $T$ still changes. Such an iteration we call degenerate.

Note that in each iteration, there can be multiple edges violating their optimality condition. There are multiple possible pivot rules that determine which edge enters $T$ in this case. In our analysis we use the (widely used in practice) pivot rule that selects as the entering edge, from all edges violating their optimality condition, the edge for which the absolute value of its reduced cost $|c^{\pi}(e)|$ is maximum. In case multiple edges in $C$ are candidates to be the leaving edge, we choose the one that is most convenient for our analysis.

If a strongly feasible spanning tree structure~\cite{AhujaNetworkFlow} is used, it can be shown that the number of iterations that the NS algorithm needs is finite. However, Zadeh~\cite{Zadeh} showed that there exist instances for which the NS algorithm (with the pivot rule stated above) needs an exponential number of iterations. Orlin~\cite{DBLP:journals/mp/Orlin97} developed a strongly polynomial version of the NS algorithm, which uses cost-scaling. However, this algorithm is rarely used in practice and we will not consider it in the rest of our paper. For a more elaborate discussion of the NS algorithm we refer to Ahuja et al.~\cite{AhujaNetworkFlow}.

\section{Upper Bound for the MMCC Algorithm} \label{UBMMCC}

In this section we show an upper bound of $O(mn^2\log(n)\log(\phi))$ for the expected number of iterations that the MMCC algorithm needs starting from the initial residual network $G_{\tilde{f}}$ for the feasible starting flow $\tilde{f}$ for flow network $G=(V,E)$. Note that we assumed in Section~\ref{IntroMCF} that $G$ is simple and that $E$ does not contain a pair $(u,v)$ and $(v,u)$ of reverse edges. This implies that for each pair of nodes $u,v\in V$, there is always at most one edge from $u$ to $v$ and at most one edge from $v$ to $u$ in any residual network $G_f$. We first show that the number of cycles that appears in at least one residual network $G_f$ for a feasible flow $f$ on $G$, is bounded by $(n+1)!$, where $n=|V|$.

\begin{lemma} \label{NumCycles}
The total number of cycles that appears in any residual network $G_f$ for a feasible flow $f$ on $G$, is bounded by $(n+1)!$.
\begin{proof}
First we show that the number of directed cycles of length $k$ ($2 \leq k \leq n$) is bounded by $n!$. We identify each cycle $C$ of length $k$ with a path $P$ of length $k$ starting and ending at the same arbitrarily chosen node of $C$ and then following the edges of $C$ in the direction of their orientation. Every such path $P$ can be identified with a unique cycle. The number $X$ of possible paths of length $k$ is bounded by
\begin{equation}
X\leq n\cdot (n-1) \cdot \ldots \cdot (n-k+1) \leq n! \label{eq:NumCycles1}
\end{equation}   

The first inequality of Equation~\eqref{eq:NumCycles1} follows since there are at most $n$ possible choices for the first node of the path, at most $n-1$ choices for the second node, etc.

The lemma follows by observing that the number of possible lengths of cycles in residual networks $G_f$ is bounded by $n+1$.
\end{proof}
\end{lemma}

We next show that the probability that any particular cycle has negative mean cost close to $0$ can be bounded. In the rest of this section, $\eps>0$.

\begin{lemma} \label{ProbSmallCost}
The probability that an arbitrary cycle $C$ has mean cost $\mu(C)\in [-\eps,0[$ can be bounded by $n\eps\phi$.
\begin{proof}  
We can only have $\mu(C)\in [-\eps,0[$ if $c(C)\in [-n\eps,0[$, since $C$ consists of at most $n$ edges.  We now draw the costs for all edges in $C$ except for the cost of one edge $e$. The cost of cycle $C$ depends linearly on the cost of edge $e$, with coefficient $1$ if $e$ is a forward edge in $C$ and coefficient $-1$ if $e$ is a reverse edge in $C$. Therefore, the width of the interval from which the cost of $e$ must be drawn such that $c(C)\in [-n\eps,0[$ is $n\eps$. Since the density function according to which the cost of $e$ is drawn has maximum density $\phi$, the probability that the cost of $e$ is drawn from this interval is at most $n\eps\phi$.
\end{proof}
\end{lemma}

\begin{corollary} \label{ProbSmallCostAllCycles}
The probability that there exists a cycle $C$ with $\mu(C)\in [-\eps,0[$ is at most $(n+1)!n\eps\phi$.
\begin{proof}
The corollary follows directly from Lemma~\ref{NumCycles} and Lemma~\ref{ProbSmallCost}.
\end{proof}
\end{corollary} 

\begin{lemma} \label{LinkNumItMeanCycleCosts}
If none of the residual networks $G_f$ for feasible flows $f$ on $G$ contain a cycle $C$ with $\mu(C)\in [-\eps,0[$, then the MMCC algorithm needs at most $mn\lceil\log_2(1/ \eps)\rceil$ iterations.
\begin{proof}
Assume to the contrary that none of the residual networks $G_f$ for feasible flows $f$ on $G$ contain a cycle $C$ with $\mu(C)\in [-\eps,0[$, but the MMCC algorithm needs more than $mn\lceil\log_2(1/ \eps)\rceil$ iterations. Let $\tilde{f}$ denote the starting flow found using any maximum-flow algorithm. Since all edge costs are drawn from the interval $[0,1]$, we have that $|\mu(\tilde{f})|\leq 1$. According to Theorem~\ref{LogBound}, after $mn\lceil\log_2(1/ \eps)\rceil$ iterations we have that for the current flow $\bar{f}$ holds that $|\mu(\bar{f})|\leq \eps$. Now either $\mu(\bar{f}) \geq 0$ contradicting the assumption that the MMCC algorithm needs more than $mn\lceil\log_2(1/ \eps)\rceil$ iterations, or $\mu(\bar{f}) \in [-\eps, 0[$ contradicting the assumption that none of the residual networks $G_f$ for feasible flows $f$ on $G$ contain a cycle $C$ with $\mu(C)\in [-\eps,0[$.
\end{proof}
\end{lemma}

\begin{theorem}
The expected number of iterations that the MMCC algorithm needs is at most $O(mn^2\log(n)\log(\phi))$. 
\begin{proof}
Let $T$ be the expected number of iterations that the MMCC algorithm needs. We have

\begin{align}
\expected(T) &= \sum_{t=1}^{\infty} \probab(T\geq t) \notag \\
&\leq \sum_{t=1}^{\infty} \probab(\textrm{any }G_f\textrm{ contains a cycle } C \textrm{ with } \mu(C)\in [-2^{-\lfloor (t-1)/mn\rfloor},0[\:) \label{eq:LinkIterationsCycles} \\
&\leq mn^2\lceil\log(n)\rceil\lceil\log(\phi)\rceil + \sum_{t=mn^2\lceil\log(n)\rceil\lceil\log(\phi)\rceil+1}^{\infty} (n+1)!n\phi 2^{-\lfloor (t-1)/mn\rfloor} \label{BoundProbCheapCycle} \\
&\leq mn^2\lceil\log(n)\rceil\lceil\log(\phi)\rceil + \sum_{t=0}^{\infty} 2^{-\lfloor t/mn\rfloor} \label{LinkNFactorialNLogN}\\
&= mn^2\lceil\log(n)\rceil\lceil\log(\phi)\rceil +mn \sum_{t=0}^{\infty} 2^{-t} \notag \\
&= O(mn^2\log(n)\log(\phi)) \notag 
\end{align}

Here Equation~\eqref{eq:LinkIterationsCycles} follows from Lemma~\ref{LinkNumItMeanCycleCosts}. Equation~\eqref{BoundProbCheapCycle} follows by bounding the probability for the first $mn^2\lceil\log(n)\rceil\lceil\log(\phi)\rceil$ terms of the summation by $1$ and the probability for the other terms using Corollary~\eqref{ProbSmallCostAllCycles}. Finally, Equation~\eqref{LinkNFactorialNLogN} follows from the inequality $n\log(n)>\log((n+2)!)$, which holds for $n\geq 6$. 

\end{proof}
\end{theorem}

\section{Lower Bound for the MMCC Algorithm} 
\subsection{General Lower Bound} \label{LBConstantPhi}
In this section we describe a construction that, for every $n$, every $m\in\{n,n+1,\ldots,n^2\}$, and every $\phi\leq2^n$, provides an instance with $\Theta(n)$ nodes and $\Theta(m)$ edges for which the MMCC algorithm requires $\Omega(m\log(\phi))$ iterations. For simplicity we describe the initial residual network $G$, which occurs after a flow satisfying all the budgets has been found, but before the first minimum-mean cycle has been canceled. For completeness, we will explain at the end of the description of $G$ how to choose the initial network, budgets, and starting flow such that $G$ is the first residual network.

\begin{figure}
\centering
\includegraphics[width=0.7\textwidth]{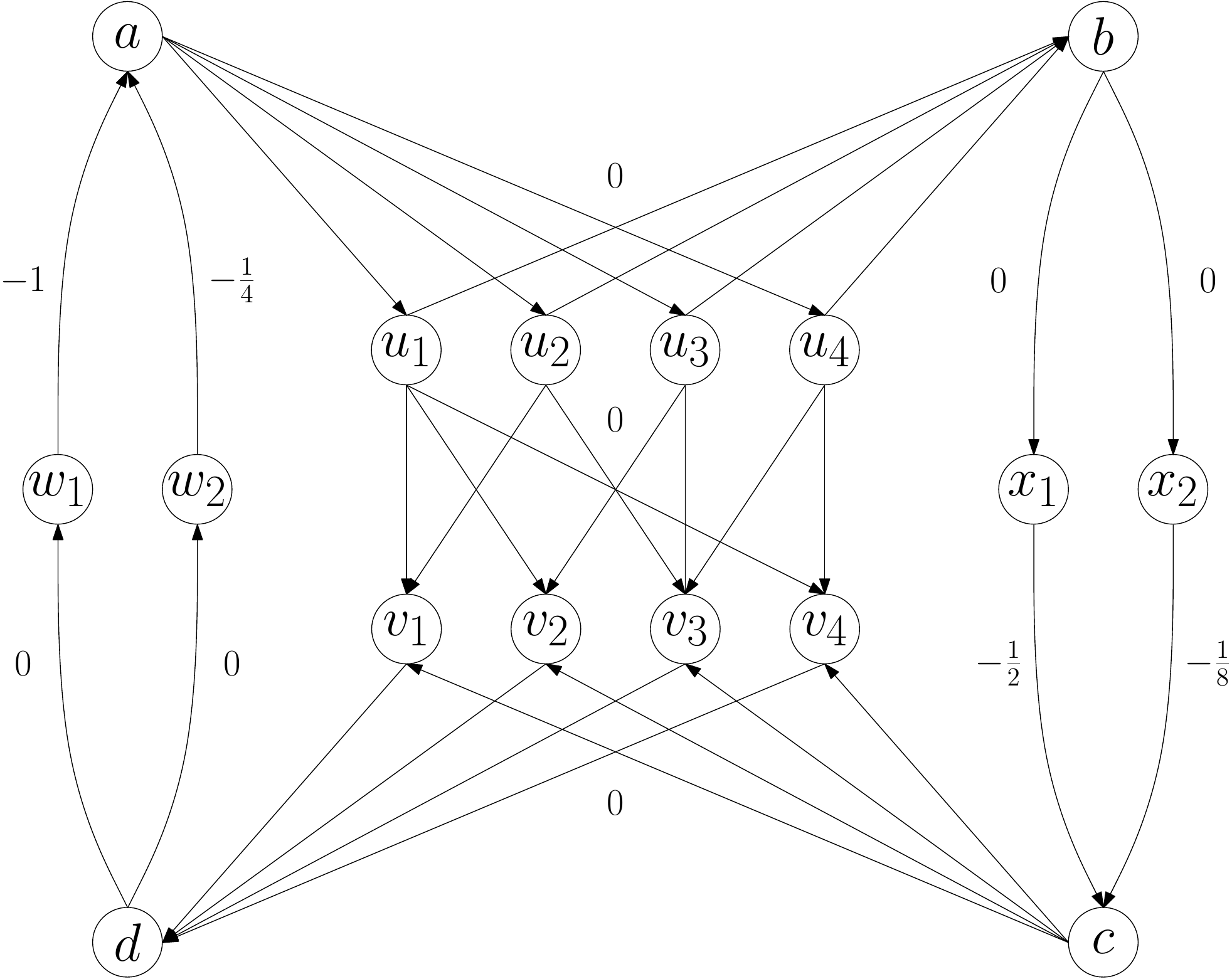}
\caption{The instance $G$ for which the MMCC algorithm needs $O(m\log(\phi))$ iterations for $n=4$, $m=9$, and $\phi=64$. Next to the edges are the approximate edge costs.}
\label{LBConstruction}
\end{figure}

We now describe how to construct $G$ given $n$, $m$, and $\phi$. In the following, we assume $\phi\geq 64$. If $\phi$ is smaller than $64$, the lower bound on the number of iterations reduces to $\Omega(m)$ and a trivial instance with $\Theta(n)$ nodes and $\Theta(m)$ edges will require $\Omega(m)$ iterations. We define $k_w=\lfloor\frac{1}{2}(\log(\phi)-4)\rfloor$ and $k_x=\lfloor\frac{1}{2}(\log(\phi)-5)\rfloor$. Note that this implies that $k_x=k_w$ or $k_x=k_w-1$. For the edge costs we define intervals from which the edge costs are drawn uniformly at random. We define $G=(\mathcal{V},\mathcal{E})$ as follows (see Figure~\ref{LBConstruction}).

\begin{itemize}
\item $\mathcal{V}=\{a,b,c,d\}\cup U \cup V \cup W \cup X$, where $U=\{u_1,\ldots,u_n\}$, $V=\{v_1,\ldots,v_n\}$, $W=\{w_1,\ldots,w_{k_w}\}$, and $X=\{x_1,\ldots,x_{k_x}\}$. 

\item $\mathcal{E}=E_{uv}\cup E_{a}\cup E_{b}\cup E_{c}\cup E_{d}\cup E_{w}\cup E_{x}$.

\item $E_{uv}$ is an arbitrary subset of $U\times V$ of cardinality $m$. Each edge $(u_i,v_j)$ has capacity $1$ and cost interval $[0,1/ \phi]$.

\item $E_a$ contains the edges $(a,u_i)$, $E_b$ contains the edges $(u_i,b)$, $E_c$ contains the edges $(c,v_i)$, and $E_d$ contains the edges $(v_i,d)$ ($i=1,\ldots,n$). All these edges have infinite capacity and cost interval $[0,1/ \phi]$.

\item $E_w$ contains the edges $(d,w_i)$ and $(w_i,a)$ ($i=1,\ldots,k_w)$. An edge $(d,w_i)$ has capacity $m$ and cost interval $[0,1/ \phi]$. An edge $(w_i,a)$ has capacity $m$ and cost interval $[-2^{2-2i}, -2^{2-2i} +1/ \phi]$.

\item $E_x$ contains the edges $(b,x_i)$ and $(x_i,c)$ ($i=1,\ldots,k_x)$. An edge $(b,x_i)$ has capacity $m$ and cost interval $[0,1/ \phi]$. An edge $(x_i,c)$ has capacity $m$ and cost interval $[-2^{1-2i}, -2^{1-2i} +1/ \phi]$.  

\end{itemize}

Note that all cost intervals have width $1/ \phi$ and therefore correspond to valid probability densities for the edge costs, since the costs are drawn uniformly at random from these intervals. The edges of the types $(w_i,a)$ and $(x_i,c)$ have a cost interval that corresponds to negative edge costs. The residual network with these negative edge costs can be obtained by having the following original instance (before computing a flow satisfying the budget requirements): All nodes, edges, costs and capacities are the same as in $G$, except that instead of the edges of type $(w_i,a)$ we have edges $(a,w_i)$ with capacity $m$ and cost interval $[2^{2-2i} -1/ \phi, 2^{2-2i} ]$ and instead of the edges of type $(x_i,c)$ we have edges $(c,x_i)$ with capacity $m$ and cost interval $[2^{1-2i} -1/ \phi, 2^{1-2i} ]$. In addition, node $a$ has budget $k_wm$, node $c$ has budget $k_xm$, the nodes of the types $w_i$ and $x_i$ have budget $-m$ and all other nodes have budget $0$. If we now choose as the initial feasible flow the flow that sends $m$ units from $a$ to each node of type $w_i$ and from $c$ to each node of type $x_i$ then we obtain the initial residual network $G$.

We now show that the MMCC algorithm needs $\Omega(m\log(\phi))$ iterations for the initial residual network $G$. First we make some basic observations. The minimum-mean cycle $C$ never contains the path $P_j = (d, w_j, a)$ if the path $P_i = (d,w_i,a)$ has positive residual capacity for some $i<j$, since the mean cost of $C$ can be improved by substituting $P_j$ by $P_i$ in $C$. Analogously, $C$ never contains the path $P_j = (b, x_j, c)$ if the path $P_i = (b,x_i,c)$ has positive residual capacity for some $i<j$. Also, since all cycles considered have mean cost strictly less than $1/ \phi$, cycles will never include more edges with cost at least $-1/ \phi$ than necessary. In addition, since the edges of type $(w_i,a)$ and $(x_i,c)$ are saturated in the order cheapest to most expensive, none of these edges will ever be included in reverse direction in the minimum-mean cycle. The above observations lead to three candidate types for the minimum-mean cycle: cycles of type $(d,w_i,a,u,v,d)$, of type $(b,x_i,c,v,u,b)$, and of type $(d,w_i,a,u,b,x_j,c,v,d)$. Here $u$ and $v$ are arbitrary nodes in $U$ and $V$, respectively. In the following series of lemmas we compare the mean costs of these cycle types. Here $u$ and $v$ are again arbitrary nodes in $U$ and $V$, possibly different for the cycles that are compared. In our computations we always assume worst-case realization of the edge costs, that is, if we want to show that a cycle $C_1$ has lower mean cost than a cycle $C_2$, we assume that all edges in $C_1$ take the highest cost in their cost interval, while all edges in $C_2$ take the lowest cost in their cost interval (an edge that appears in both $C_1$ and $C_2$ can even take its highest cost in $C_1$ and its lowest cost in $C_2$ in the analysis). 

\begin{lemma} \label{ADBC}
The cycle $C_1= (d,w_i,a,u,v,d)$ has lower mean cost than the cycle $C_2=(b,x_i,c,v,u,b)$.
\begin{proof}
Since the cycles have equal length, we can compare their total costs instead of their mean costs. We have
\begin{align*}
w(C_1)-w(C_2) &\leq \left(-2^{2-2i} +5/ \phi\right) - \left(-2^{1-2i}-1/ \phi\right) \\
&\leq -64/ \phi+6/ \phi <0
\end{align*}
Here the second inequality holds since $i\leq k_x \leq \frac{1}{2}(\log(\phi)-5)$. 
\end{proof}
\end{lemma}  
    
\begin{lemma} \label{BCADIPlus1}
The cycle $C_1=(b,x_i,c,v,u,b)$ has lower mean cost than the cycle $C_2= (d,w_{i+1},a,u,v,d)$.
\begin{proof}
Since the cycles have equal length, we can compare their total costs instead of their mean costs. We have
\begin{align*}
w(C_1)-w(C_2) &\leq \left(-2^{1-2i} +4/ \phi\right) - \left(-2^{2-2(i+1)}\right) \\
&\leq -64/ \phi+4/ \phi <0
\end{align*}
Here the second inequality holds since $i+1\leq k_w \leq \frac{1}{2}(\log(\phi)-4)$. 
\end{proof}
\end{lemma}

\begin{lemma} \label{ADCycle}
The cycle $C_1=(d,w_i,a,u,v,d)$ has lower mean cost than the cycle $C_2= (d,w_i,a,u,b,x_i,c,v,d)$.
\begin{proof}
We have
\begin{align*}
w(C_1)/ |C_1|-w(C_2)/ |C_2| &\leq \left(-2^{2-2i} +\frac{5}{\phi}\right)/5 - \left(-2^{2-2i} -2^{1-2i}\right)/8 \\
&\leq \frac{-8}{5\phi}+\frac{1}{\phi} <0
\end{align*}
Here the second inequality holds since $i\leq k_x \leq \frac{1}{2}(\log(\phi)-5)$. 
\end{proof}
\end{lemma}

\begin{lemma} \label{BCCycle}
The cycle $C_1=(b,x_i,c,v,u,b)$ has lower mean cost than the cycle $C_2= (b,x_i,c,v,d,w_{i+1},a,u,b)$.
\begin{proof}
We have
\begin{align*}
w(C_1)/ |C_1|-w(C_2)/ |C_2| &\leq \left(-2^{1-2i} +\frac{4}{\phi}\right)/5 - \left(-2^{1-2i} -2^{2-2(i+1)}\right)/8 \\
&\leq \frac{-8}{5\phi}+\frac{4}{5\phi} <0
\end{align*}
Here the second inequality holds since $i+1\leq k_w \leq \frac{1}{2}(\log(\phi)-4)$. 
\end{proof}
\end{lemma}   

The above observations and lemmas allow us to determine the number of iterations that the MMCC algorithm needs for residual network $G$. 
    
\begin{theorem} \label{LowerBoundG}
The MMCC algorithm needs $m(k_w+k_x)$ iterations for residual network $G$, independent of the realization of the edge costs. 
\begin{proof}
For the first iteration only cycles of the types $(d,w_i,a,u,v,d)$ and $(d,w_i,a,u,b,x_i,c,v,d)$ are available. According to Lemma~\ref{ADCycle}, all cycles of type $(d,w_i,a,u,v,d)$ have lower mean costs than all cycles of type $(d,w_i,a,u,b,x_i,c,v,d)$ and therefore the first iteration will augment along a cycle of type $(d,w_i,a,u,v,d)$. After the first iteration, an edge from $V$ to $U$ will become available, and therefore a cycle of the type $(b,x_i,c,v,u,b)$. According to Lemma~\ref{ADBC} this cycle has higher mean cost than cycles of the type $(d,w_i,a,u,v,d)$ however, and therefore the first $m$ iterations will be of the type $(d,w_i,a,u,v,d)$. 

After the first $m$ iterations, the edge $(d,w_1)$, the edge $(w_1,a)$, and all edges in $E_{uv}$ will be saturated. The available cycle types are now $(b,x_i,c,v,u,b)$ and $(b,x_i,c,v,d,w_{i+1},a,u,b)$. According to Lemma~\ref{BCCycle}, all cycles of type $(b,x_i,c,v,u,b)$ have lower mean cost than all cycles of type $(b,x_i,c,v,d,w_{i+1},a,u,b)$. The next iteration will therefore augment along a cycle of type $(b,x_i,c,v,u,b)$. After this iteration, an edge from $U$ to $V$ becomes available and therefore a cycle of type $(d,w_{i+1},a,u,v,d)$, but according to Lemma~\ref{BCADIPlus1} all cycles of type $(b,x_i,c,v,u,b)$ have lower mean cost than cycles of type $(d,w_{i+1},a,u,v,d)$ and therefore in iterations $m+1,\ldots,2m$ the MMCC algorithm augments along cycles of type $(b,x_i,c,v,u,b)$.

After $2m$ iterations, we obtain $G$, except that now edges $(d,w_1)$, $(w_1,a)$, $(b,x_1)$, and $(x_1,c)$ are saturated and that there is some flow on the infinite capacity edges of the types $(a,u_i)$, $(u_i,b)$, $(c,v_i)$, and $(v_i,d)$. The MMCC algorithm will keep augmenting among $m$ cycles of type $(d,w_i,a,u,v,d)$ followed by $m$ cycles of type $(b,x_i,c,v,u,b)$ until all edges of types $(w_i,a)$ and $(x_i,c)$ are saturated and no negative cost cycles remain. The total number of iterations the MMCC algorithm needs is therefore $m(k_w+k_x)$.
\end{proof}
\end{theorem}

The instance $G$ and Theorem~\ref{LowerBoundG} allow us to state a lower bound on the number of iterations that the MMCC algorithm needs in the smoothed setting.

\begin{theorem}
For every $n$, every $m\in\{n,n+1,\ldots,n^2\}$, and every $\phi\leq2^n$, there exists an instance with $\Theta(n)$ nodes and $\Theta(m)$ edges for which the MMCC algorithm requires $\Omega(m\log(\phi))$ iterations, independent of the realization of the edge costs.
\begin{proof}
Follows directly from the instance $G$, Theorem~\ref{LowerBoundG} and the definition of $k_w$ and $k_x$.
\end{proof}
\end{theorem}

\subsection{Lower Bound for $\phi$ Dependent on $n$} \label{LBBigPhi}
In Section~\ref{LBConstantPhi} we considered the setting where $\phi$ does not depend on $n$. In this setting we showed that the MMCC algorithm needs $\Omega(m\log(\phi))$ iterations. We can improve the lower bound if $\phi$ is much larger than $n$. In this section we consider the case where $\phi=\Omega(n^2)$. In particular, we show that for every $n\geq 4$ and every $m\in\{n,\ldots,n^2\}$ there exists an instance with $\Theta(n)$ nodes, $\Theta(m)$ edges, and $\phi=\Theta(n^2)$ for which the MMCC algorithm needs $\Omega(mn)$ iterations.

\begin{figure}
\centering
\includegraphics[width=0.7\textwidth]{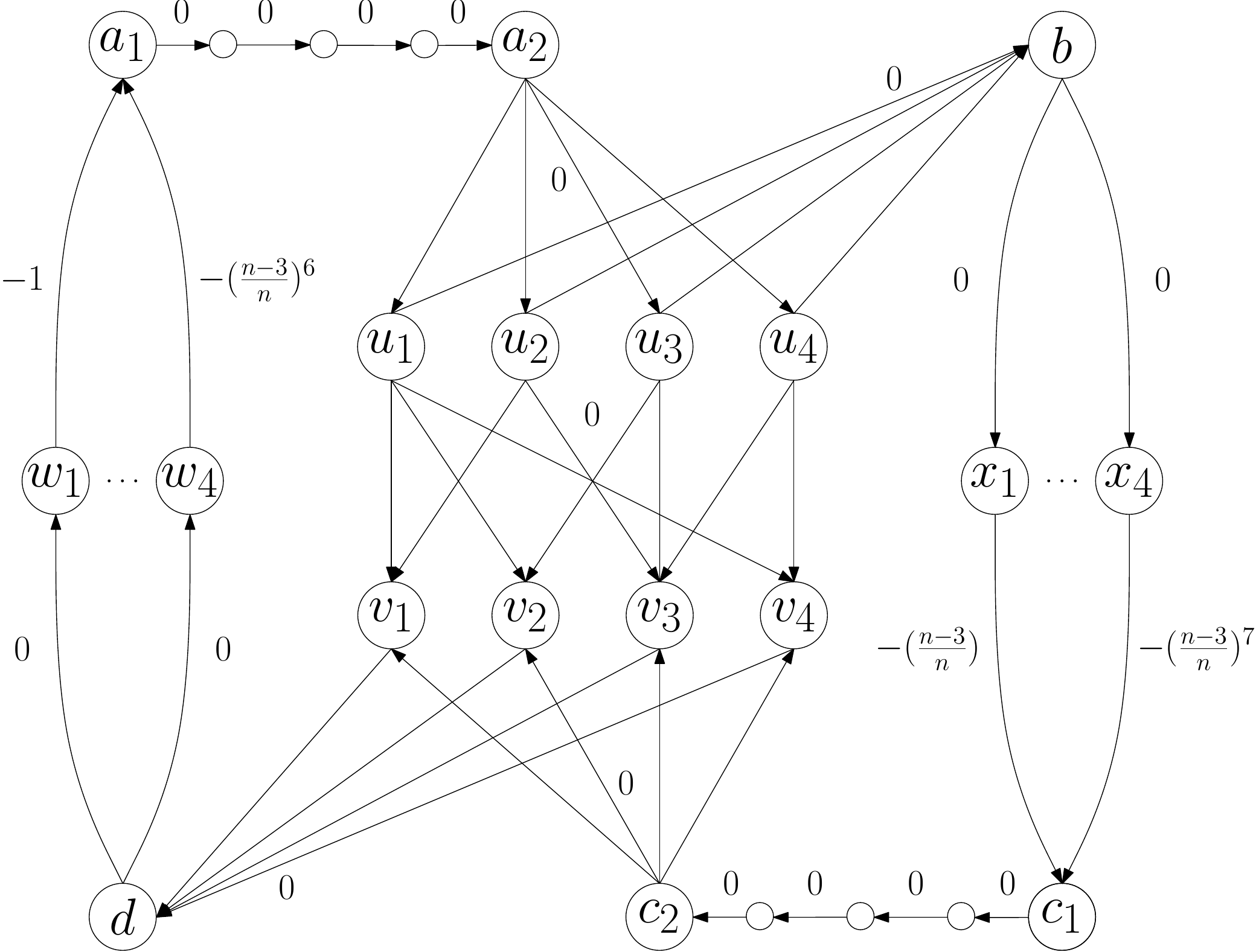}
\caption{The instance $G$ for which the MMCC algorithm needs $\Omega(mn)$ iterations for $n=4$, $m=9$, and $\phi=400000n^2$. Next to the edges are the approximate edge costs.}
\label{LBConstructionBigPhi}
\end{figure}

The initial residual network $H$ that we use to show our bound is very similar to the initial residual network $G$ that was used to show the bound in Section~\ref{LBConstantPhi}. Below we describe the differences (see Figure~\ref{LBConstructionBigPhi} for an illustration). We set $\phi=400000n^2$. The constant of $400000$ is large, but for the sake of readability and ease of calculations we did not try to optimize it.

\begin{itemize}
\item The node set $W$ now consists of $n$ nodes $\{w_1,\ldots,w_n\}$ and the node set $X$ now consists of $n$ nodes $\{x_1,\ldots,x_n\}$.
\item Node $a$ is split into two nodes $a_1$ and $a_2$. From node $a_1$ to $a_2$ there is a directed path consisting of $n$ edges, all with infinite capacity and cost interval $[0,1/ \phi]$. Edges $(a,u_i)$ are replaced by edges $(a_2,u_i)$ with infinite capacity and cost interval $[0,1/ \phi]$. Edges $(w_i,a)$ are replaced by edges $(w_i,a_1)$ with capacity $m$ and cost interval $[{-(\frac{n-3}{n})}^{2i-2},{-(\frac{n-3}{n})}^{2i-2}+\frac{1}{\phi}]$.       
\item Node $c$ is split into two nodes $c_1$ and $c_2$. From node $c_1$ to $c_2$ there is a directed path consisting of $n$ edges, all with infinite capacity and cost interval $[0,1/ \phi]$. Edges $(c,v_i)$ are replaced by edges $(c_2,v_i)$ with infinite capacity and cost interval $[0,1/ \phi]$. Edges $(x_i,c)$ are replaced by edges $(x_i,c_1)$ with capacity $m$ and cost interval $[{-(\frac{n-3}{n})}^{2i-1},{-(\frac{n-3}{n})}^{2i-1}+\frac{1}{\phi}]$.
\end{itemize}   

Note that this is a valid choice of cost intervals for the edges $(w_i,a_1)$ and $(x_i,c_1)$ and that they all have negative costs, since $(x_n,c_1)$ is the most expensive of them and we have

\begin{equation}
{-\left(\frac{n-3}{n}\right)}^{2n-1}+\frac{1}{\phi} \leq {-\left(1-\frac{3}{n}\right)}^{2n} + \frac{1}{400000n^2} \leq -(e^{-6})^2 +\frac{1}{6400000} <0\text{.}
\end{equation} 

As in Section~\ref{LBConstantPhi}, there are three candidate types for the minimum-mean cost cycle: cycles of type $(d,w,a,u,v,d)$, cycles of type $(b,x,c,v,u,b)$, and cycles of type $(d,w,a,u,b,x,c,v,d)$. Again we assume worst-case realizations of the edge costs and compare the mean costs of cycles of the different types in a series of lemmas.

\begin{lemma} \label{ADBCBigPhi}
The cycle $C_1= (d,w_i,a,u,v,d)$ has lower mean cost than the cycle $C_2=(b,x_i,c,v,u,b)$.
\begin{proof}
Since the cycles have equal length, we can compare their total costs instead of their mean costs. We have
\begin{align*}
w(C_1)-w(C_2) &\leq \left(-\left(\frac{n-3}{n}\right)^{2i-2}+\frac{n+5}{\phi}\right) - \left(-\left(\frac{n-3}{n}\right)^{2i-1}-\frac{1}{\phi}\right) \\
&\leq -\frac{3e^{-12}}{n}+\frac{n+6}{\phi} <0
\end{align*}
Here the second inequality holds since $i\leq n$ and $-\left(\frac{n-3}{n}\right)^{2n}\leq -e^{-12}$ for $n\geq 4$. 
\end{proof}
\end{lemma}  
    
\begin{lemma} \label{BCADIPlus1BigPhi}
The cycle $C_1=(b,x_i,c,v,u,b)$ has lower mean cost than the cycle $C_2= (d,w_{i+1},a,u,v,d)$.
\begin{proof}
Since the cycles have equal length, we can compare their total costs instead of their mean costs. We have
\begin{align*}
w(C_1)-w(C_2) &\leq \left(-\left(\frac{n-3}{n}\right)^{2i-1}+\frac{n+4}{\phi}\right) - \left(-\left(\frac{n-3}{n}\right)^{2(i+1)-2}\right) \\
&\leq -\frac{3e^{-12}}{n}+\frac{n+4}{\phi} <0
\end{align*}
Here the second inequality holds since $i+1\leq n$ and $-\left(\frac{n-3}{n}\right)^{2n}\leq -e^{-12}$ for $n\geq 4$. 
\end{proof}
\end{lemma}

\begin{lemma} \label{ADCycleBigPhi}
The cycle $C_1=(d,w_i,a,u,v,d)$ has lower mean cost than the cycle $C_2= (d,w_i,a,u,b,x_i,c,v,d)$.
\begin{proof}
We have
\begin{align*}
w(C_1)/ |C_1|-w(C_2)/ |C_2| &\leq \frac{\left(-\left(\frac{n-3}{n}\right)^{2i-2}+\frac{n+5}{\phi}\right)}{n+5} - \frac{\left(-\left(\frac{n-3}{n}\right)^{2i-2}-\left(\frac{n-3}{n}\right)^{2i-1}\right)}{2n+8} \\
&\leq -e^{-12}\left(\frac{n+15}{(n+5)(2n+8)n}\right)+\frac{1}{\phi} <0
\end{align*}
Here the second inequality holds since $i\leq n$ and $-\left(\frac{n-3}{n}\right)^{2n}\leq -e^{-12}$ for $n\geq 4$. 
\end{proof}
\end{lemma}

\begin{lemma} \label{BCCycleBigPhi}
The cycle $C_1=(b,x_i,c,v,u,b)$ has lower mean cost than the cycle $C_2= (b,x_i,c,v,d,w_{i+1},a,u,b)$.
\begin{proof}
We have
\begin{align*}
w(C_1)/ |C_1|-w(C_2)/ |C_2| &\leq \frac{\left(-\left(\frac{n-3}{n}\right)^{2i-1}+\frac{n+4}{\phi}\right)}{n+5} - \frac{\left(-\left(\frac{n-3}{n}\right)^{2i-1}-\left(\frac{n-3}{n}\right)^{2(i+1)-2}\right)}{2n+8} \\
&\leq -e^{-12}\left(\frac{n+15}{(n+5)(2n+8)n}\right)+\frac{n+4}{(n+5)\phi} <0
\end{align*}
Here the second inequality holds since $i+1\leq n$ and $-\left(\frac{n-3}{n}\right)^{2n}\leq -e^{-12}$ for $n\geq 4$. 
\end{proof}
\end{lemma}

The above lemmas allow us to determine the number of iterations that the MMCC algorithm needs for initial residual network $H$. 

\begin{theorem} \label{LowerBoundH}
The MMCC algorithm needs $2mn$ iterations for initial residual network $H$, independent of the realization of the edge costs.
\begin{proof}
The proof is similar to the proof of Theorem~\ref{LowerBoundG}. Since cycles of type $(d,w_i,a,u,v,d)$ have lower mean cost than both cycles of type $(b,x_i,c,v,u,b)$ and of type $(d,w_i,a,u,b,x_i,c,v,d)$ according to Lemma~\ref{ADBCBigPhi} and Lemma~\ref{ADCycleBigPhi}, the first $m$ iterations will augment flow along cycles of type $(d,w_i,a,u,v,d)$. After these $m$ iterations, edges $(d,w_1)$ and $(w_1,a_1)$ are saturated and the edges in $E_{uv}$ have positive residual capacity only in the direction from $V$ to $U$. 

Since cycles of type $(b,x_i,c,v,u,b)$ have lower mean cost than both cycles of type $(d,w_{i+1},a,u,v,d)$ and cycles of type $(b,x_i,c,v,d,w_{i+1},a,u,b)$ according to Lemma~\ref{BCADIPlus1BigPhi} and Lemma~\ref{BCCycleBigPhi}, the next $m$ iterations will augment flow along cycles of type $(b,x_i,c,v,u,b)$. After the first $2m$ iterations, the residual network is the same as $H$, except that edges $(d,w_1)$, $(w_1,a_1)$, $(b,x_1)$, and $(x_1,c_1)$ are saturated and there is some flow on several edges of infinite capacity. The MMCC algorithm will keep augmenting along $m$ cycles of type $(d,w_i,a,u,v,d)$ followed by $m$ cycles of type $(b,x_i,c,v,u,b)$, until all edges of type $(w_i,a_1)$ and type $(x_i,c_1)$ are saturated. At this point no negative cycles remain in the residual network and the MMCC algorithm terminates after $2mn$ iterations.
\end{proof}
\end{theorem}  

Initial residual network $H$ and Theorem~\ref{LowerBoundH} allow us to state a lower bound for the number of iterations that the MMCC Algorithm needs in the smoothed setting for large $\phi$.

\begin{theorem}
For every $n\geq 4$ and every $m\in\{n,n+1,\ldots,n^2\}$, there exists an instance with $\Theta(n)$ nodes and $\Theta(m)$ edges, and $\phi=\Theta(n^2)$, for which the MMCC algorithm requires $\Omega(mn)$ iterations, independent of the realization of the edge costs.
\begin{proof}
Follows directly from the instance $H$ and Theorem~\ref{LowerBoundH}.
\end{proof}
\end{theorem}

\section{Lower bound for the Network Simplex Algorithm} \label{LBNS}
In this section we provide a lower bound of $\Omega(m\cdot \min\{n,\phi\}\cdot \phi)$ for the number of iterations that the Network Simplex (NS) algorithm requires in the setting of smoothed analysis. The instance of the minimum-cost flow problem that we use to show this lower bound is very similar to the instance used by Brunsch et al.~\cite{BrunschSSP} to show a lower bound on the number of iterations that the Successive Shortest Path algorithm needs in the smoothed setting. The differences are that they scaled their edge costs by a factor of $\phi$, which we do not, that we add an extra path from node $s$ to node $t$, and that the budgets of the nodes are defined slightly differently. For completeness, we describe the construction of the (slightly adapted) instance by Brunsch et al.\ below. For a more elaborate description of the instance, we refer to the original paper. In the following we assume that all paths from $s$ to $t$ have pairwise different costs, which holds with probability 1, since the edge costs are drawn from continuous probability distributions. 

For given positive integers~$n$, $m \in \{ n, \ldots, n^2 \}$, and $\phi \leq 2^n$ let $k = \lfloor{\log \phi}\rfloor - 5 = O(n)$ and $M = \min \{ n, 2^{\lfloor{\log_2 \phi}\rfloor}/4-2 \} = \Theta(\min \{ n, \phi \})$. Like Brunsch et al.\ we assume that $\phi\geq 64$, which implies $k,M\geq 1$. We construct a flow network with $\Theta(n)$ nodes, $\Theta(m)$ edges, and smoothing parameter $\phi$ for which the NS algorithm needs $\Theta(m\cdot \min\{n,\phi\}\cdot \phi)$ iterations. The construction of the flow network $G$ that we use to show our lower bound consists of three steps. First we define a flow network $G_1$. Next we describe how to obtain flow network $G_{i+1}$ from flow network $G_{i}$. Finally, we construct $G$ using $G_k$. As before, we give an interval of width at least $1/ \phi$ for each edge from which its costs are drawn uniformly at random. 

\begin{figure}
\centering
\includegraphics[width=0.7\textwidth]{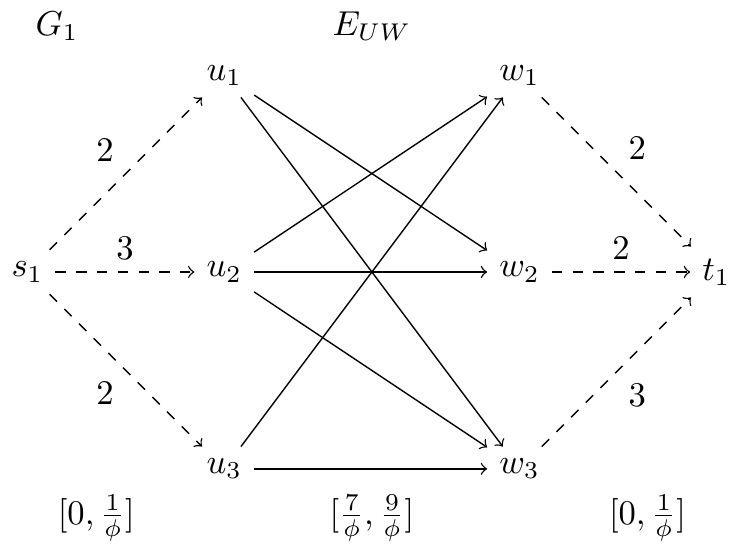}
\caption{Example for $G_1$ with $n=3$ and $m = 7$ with capacities different from 1 shown next to the edges and the cost intervals shown below each edge set. Dashed edges are in the initial spanning tree for the network simplex algorithm, while solid edges are not.}
\label{G1}
\end{figure}

\paragraph{\boldmath Construction of~$G_1$.}
For the first step, consider two sets $U = \{ u_1, \ldots, u_n \}$ and $W = \{ w_1, \ldots, w_n \}$ of~$n$ nodes and an arbitrary set $E_{UW} \subseteq U \times W$ containing exactly $|E_{UW}| = m$ edges. The initial flow network~$G_1$ is defined as $G_1 = (V_1, E_1)$ for $V_1 = U \cup W \cup \{ s_1, t_1 \}$ and
\[
  E_1 = (\{s_1\} \times U) \cup E_{UW} \cup (W \times \{t_1\}).
\]
The edges~$e$ from~$E_{UW}$ have capacity~$1$ and costs from the interval $I_e = [\frac{7}{\phi}, \frac{9}{\phi}]$. 
The edges $(s_1,u_i), u_i \in U$ have capacity equal to the out-degree of~$u_i$, 
the edges $(w_j,t_1), w_j \in W$ have capacity equal to the in-degree of $w_j$ and both have costs from the interval $I_e = [0, \frac{1}{\phi}]$ (see Figure~\ref{G1}).

\begin{figure}
\centering
\includegraphics[width=0.7\textwidth]{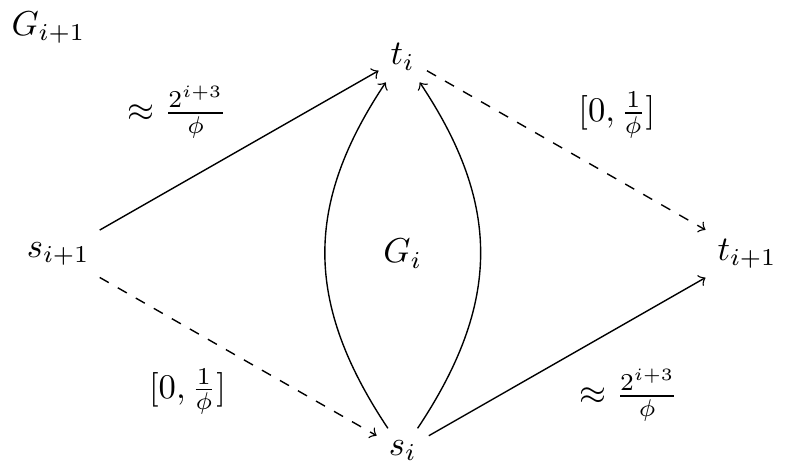}
\caption{$G_{i+1}$ with $G_i$ as sub-graph with edge costs next to the edges. Dashed edges are in the initial spanning tree for the network simplex algorithm, while solid edges are not.}
\label{GIPlus1}
\end{figure}

\paragraph{\boldmath Construction of~$G_{i+1}$ from~$G_i$.}
Now we describe the second step of our construction. Given a flow network $G_i = (V_i, E_i)$, we define $G_{i+1} = (V_{i+1}, E_{i+1})$, where $V_{i+1} = V_i \cup \{ s_{i+1}, t_{i+1} \}$ and
\[
  E_{i+1} = E_{i} \cup (\{s_{i+1}\} \times \{ s_i, t_i \}) \cup (\{ s_i, t_i \} \times \{t_{i+1}\}).
\]
Let~$N_i$ be the value of the maximum $s_i$-$t_i$ flow in the graph~$G_i$. The new edges $e \in \{ (s_{i+1}, s_i), (t_i, t_{i+1}) \}$ have capacity $u(e) = N_i$ and costs from the interval $I_e = [0, \frac{1}{\phi}]$. The new edges $e \in \{ (s_{i+1}, t_i), (s_i, t_{i+1}) \}$ also have capacity $u(e) = N_i$, but costs from the interval $I_e = [\frac{2^{i+3}-1}{\phi}, \frac{2^{i+3}+1}{\phi}]$ (see Figure~\ref{GIPlus1}).

\begin{figure}
\centering
\includegraphics[width=1.0\textwidth]{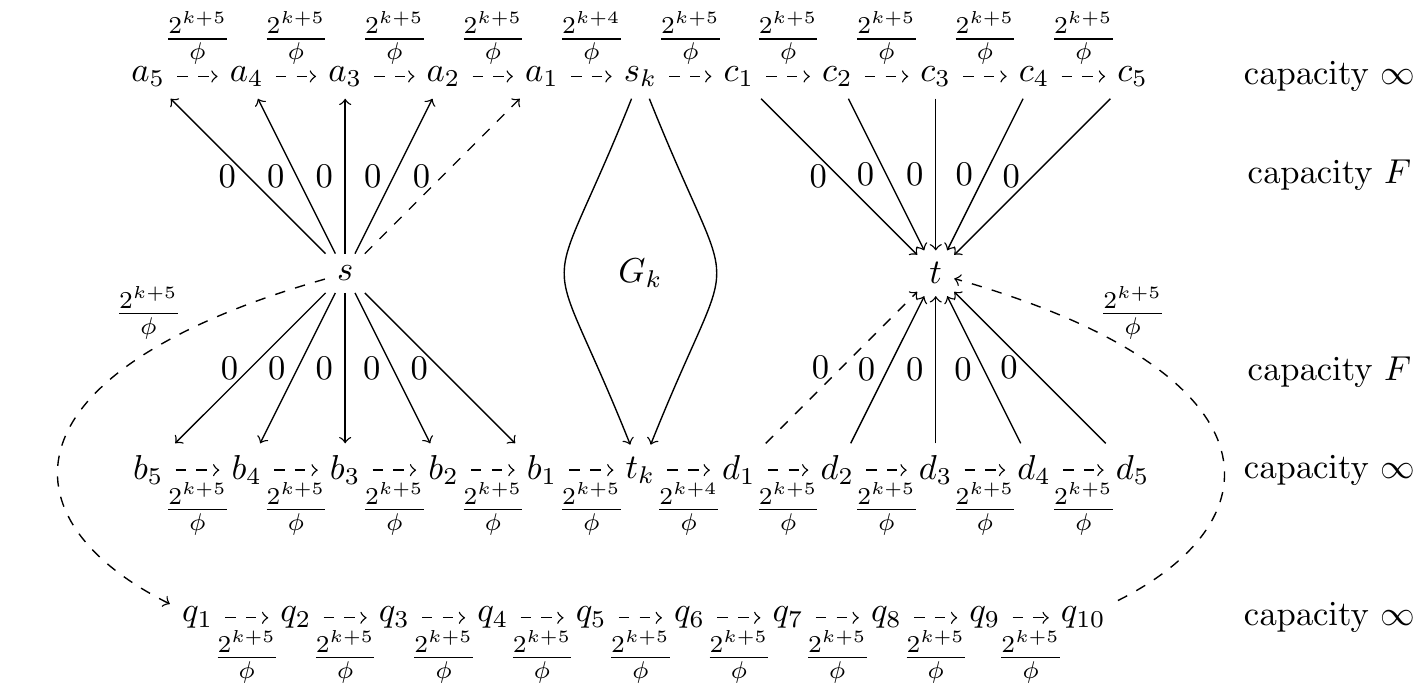}
\caption{$G$ with $G_k$ as sub-graph with approximate edge costs next to the edges. Dashed edges are in the initial spanning tree for the network simplex algorithm, while solid edges are not.}
\label{GNetworkSimplex}
\end{figure}

\paragraph{\boldmath Construction of~$G$ from~$G_k$.}
Let $F$ be the value of a maximum $s_k$-$t_k$ flow in $G_k$. We will now use $G_k$ to define $G=(V,E)$ as follows
(see also Figure~\ref{GNetworkSimplex}).
\begin{itemize}
\setlength{\itemsep}{0cm}
\item $V := V_k \cup A \cup B \cup C \cup D \cup Q \cup \{s,t\}$, with $A := \{a_1,a_2,\dots,a_M\}$, $B := \{b_1,b_2,\dots,b_M\}$, $C := \{c_1,c_2,\dots,c_M\}$, $D := \{d_1,d_2,\dots,d_M\}$, and $Q := \{q_1,q_2,\dots,q_{2M}\}$. \newline
\item $E:= E_k \cup E_a \cup E_b \cup E_c \cup E_d \cup E_q$.\newline
\item $E_a$ contains the edges $(a_i,a_{i-1})$, $i \in \{2,\dots,M\}$, with cost interval $[\frac{2^{k+5}-1}{\phi},\frac{2^{k+5}}{\phi}]$ and infinite capacity;
$(s,a_i)$, $i \in \{1,\dots,M\}$, with cost interval $[0,\frac{1}{\phi}]$ and capacity $F$; and $(a_1,s_k)$ with cost interval $[\frac{2^{k+4}-1}{\phi},\frac{2^{k+4}}{\phi}]$ and infinite capacity.\newline
\item $E_b$ contains the edges $(b_i,b_{i-1})$, $i \in \{2,\dots,M\}$, with cost interval $[\frac{2^{k+5}-1}{\phi},\frac{2^{k+5}}{\phi}]$ and infinite capacity;
$(s,b_i)$, $i \in \{1,\dots,M\}$, with cost interval $[0,\frac{1}{\phi}]$ and capacity $F$; and $(b_1,t_k)$ with cost interval $[\frac{2^{k+5}-1}{\phi},\frac{2^{k+5}}{\phi}]$ and infinite capacity.\newline
\item $E_c$ contains the edges $(c_{i-1},c_i)$, $i \in \{2,\dots,M\}$, with cost interval $[\frac{2^{k+5}-1}{\phi},\frac{2^{k+5}}{\phi}]$ and infinite capacity;
$(c_i,t)$, $i \in \{1,\dots,M\}$, with cost interval $[0,\frac{1}{\phi}]$ and capacity $F$; and $(s_k,c_1)$ with cost interval $[\frac{2^{k+5}-1}{\phi},\frac{2^{k+5}}{\phi}]$ and infinite capacity.\newline
\item $E_d$ contains the edges $(d_{i-1},d_i)$, $i \in \{2,\dots,M\}$, with cost interval $[\frac{2^{k+5}-1}{\phi},\frac{2^{k+5}}{\phi}]$ and infinite capacity;
$(d_i,t)$, $i \in \{1,\dots,M\}$, with cost interval $[0,\frac{1}{\phi}]$ and capacity $F$; and $(t_k,d_1)$ with cost interval $[\frac{2^{k+4}-1}{\phi},\frac{2^{k+4}}{\phi}]$ and infinite capacity.\newline
\item $E_q$ contains the edges $(q_{i-1},q_i)$, $i \in \{2,\dots,2M\}$, with cost interval $[\frac{2^{k+5}-1}{\phi},\frac{2^{k+5}}{\phi}]$ and infinite capacity;
$(s,q_1)$, with cost interval $[\frac{2^{k+5}-1}{\phi},\frac{2^{k+5}}{\phi}]$ and infinite capacity; and $(q_{2M},t)$ with cost interval $[\frac{2^{k+5}-1}{\phi},\frac{2^{k+5}}{\phi}]$ and infinite capacity.
\end{itemize}

The budgets of all nodes are $0$, except for node $s$ and $t$, which have budgets $b(s)=2MF$ and $b(t)=-2MF$. We choose as the initial spanning tree $T$ for the NS algorithm the edges 
\begin{itemize}
\item $(s_1,u_i)$ ($i=1,\ldots,n$) and $(w_i,t_1)$ ($i=1,\ldots,n$),
\item $(s_{i+1},s_i)$ ($i=1,\ldots,k-1$) and $(t_{i},t_{i+1})$ ($i=1,\ldots,k-1$),
\item $(s,a_1)$, $(a_i,a_{i-1})$ ($i =2,\ldots,M$), and $(a_1,s_k)$,
\item $(s_k,c_1)$ and $(c_{i-1},c_{i})$ ($i =2,\ldots,M$),
\item $(b_i,b_{i-1})$ ($i =2,\ldots,M$) and $(b_1,t_k)$, 
\item $(d_1,t)$, $(t_k,d_1)$, and $(d_{i-1},d_{i})$ ($i =2,\ldots,M$),
\item $(s,q_1)$, $(q_i,q_{i+1})$ ($i =1,\ldots,2M-1$), and $(q_{2M},t)$.
\end{itemize}
(see Figure~\ref{G1}, Figure~\ref{GIPlus1}, and Figure~\ref{GNetworkSimplex}). In addition, $\tilde{L}=E\backslash T$ and $\tilde{U}=\emptyset$. Spanning tree structure $(T,\tilde{L},\tilde{U})$ corresponds to the flow that sends $2MF$ units of flow on the path $(s,q_1,\ldots,q_{2M},t)$ and does not send any flow on other edges.

To prove our lower bound on the number of iterations that the NS algorithm needs for flow network $G$, we link the iterations of the NS algorithm to the iterations of the Successive Shortest Path (SSP) algorithm for flow network $H$, where $H$ is obtained from $G$ by removing all nodes $q_i$ ($i=1,\ldots,2M$) and all their incident edges. As shown by Brunsch et al.~\cite[Theorem 23]{BrunschSSP}, the SSP algorithm needs $\Omega(m\cdot \min\{n,\phi\}\cdot \phi)$ iterations for $H$. We show that each non-degenerate iteration of the NS algorithm on $G$ corresponds with an iteration of the SSP algorithm on $H$ and that therefore the NS algorithm needs $\Omega(m\cdot \min\{n,\phi\}\cdot \phi)$ iterations for flow network $G$ as well. In our analysis we use many results on the costs of paths in $H$ by Brunsch et al.~\cite{BrunschSSP}. We will not prove these results again, but refer to the original paper. 

We now first observe that path $(s,q_1,\ldots,q_{2M},t)$ is the most expensive path from $s$ to $t$ in $G$. Since paths from $s_k$ to $t_k$ have cost less than $2^{k+3}$~\cite[Lemma 21]{BrunschSSP}, the most expensive $s-t$ paths that do not use the nodes $q_i$ are $(s,a_M,\ldots,c_M,t)$ and $(s,b_M,\ldots,d_M,t)$. Both those paths have the same distribution for their costs. If we arbitrarily choose path $(s,a_M,\ldots,c_M,t)$ and compare its cost with the cost of path $(s,q_1,\ldots,q_{2M},t)$, assuming worst case edge cost realizations, we have
\begin{align*}
& \qquad c(s,q_1,\ldots,q_{2M},t)- c(s,a_M,\ldots,c_M,t) \\
&\geq (2M+1)\left(\frac{2^{k+5}-1}{\phi}\right)-\left(\frac{(2M-1)2^{k+5}+2^{k+4}+2}{\phi}\right) \\
&=\frac{3\cdot2^{k+4}-2M-3}{\phi} >0,
\end{align*}  
by the definition of $k$ and $M$ and $\phi\geq 64$, which shows that path $(s,q_1,\ldots,q_{2M},t)$ is the most expensive path from $s$ to $t$ in $G$.

Since $(s,q_1,\ldots,q_{2M},t)$ is the most expensive $s-t$ path, we can obtain a negative cycle $C$ by combining the reverse edges of $(s,q_1,\ldots,q_{2M},t)$ with the edges of another $s-t$ path $P$. Clearly, the cheaper $P$ is, the cheaper is $C$. In addition, if all edges of $C$ except for one edge $(i,j)$ are in the current spanning tree $\tilde{T}$, then edge $(i,j)$ has reduced cost equal to the cost of $C$, since all edges in $\tilde{T}$ have reduced cost $0$. Using these two observations, we can conclude that as long as all negative cycles that can be formed using edges in the current spanning tree $\tilde{T}$ plus one edge outside of $\tilde{T}$ that has positive residual capacity in the current residual network consist of path $(t,q_{2M},\ldots,q_{1},s)$ plus an $s-t$ path $P$, then the NS algorithm with the most negative edge pivot rule will pivot on the edge that together with the edges in $\tilde{T}$ forms the cheapest $s-t$ path.

Brunsch et al.~\cite{BrunschSSP} showed that the SSP algorithm encounters $2MF$ paths on $H$. Let $P_1,\ldots,P_{2MF}$ be the paths encountered on $H$ by the SSP algorithm, ordered from cheapest to most expensive. We refer the reader to the paper of Brunsch et al.\ for a description of these paths. In the following we show that in the $i^{th}$ non-degenerate iteration of the NS algorithm on $G$, flow is sent along cycle $(t,q_{2M},\ldots,q_1,s)\cup P_i$. We will not provide all the calculations needed to compare the reduced cost of the candidate edges for addition to the spanning tree in each iteration, but it can be checked by tedious computation that the claimed edge is indeed the one with lowest reduced cost. If flow is sent over one of the edges in $E_{UW}$, that edge is always a candidate leaving edge, since edges in $E_{UW}$ have capacity 1 and all other capacities are integral. For simplicity we assume that in cases where multiple edges in the cycle become saturated simultaneously by the flow augmentation, it is always the edge from $E_{UW}$ that leaves the spanning tree. 

In the first iteration of the NS algorithm on $G$, all edges in $E_{UW}$ have negative reduced cost and positive residual capacity, and the edge $(u_i,w_j)$ that together with the starting spanning tree $T$ contains path $P_1$ is added to $T$, flow is augmented along cycle $(t,q_{2M},\ldots,q_1,s)\cup P_1$, and $(u_i,w_j)$ becomes saturated and leaves $T$ again. For the second iteration, edge $(u_i,w_j)$ is saturated and therefore the edge that together with $T$ contains $P_2$ will be added to $T$. This will continue for the first $m$ iterations. 

At this point, the cheapest $s-t$ path using edges in $T$ plus one edge with positive residual capacity is the path that is obtained by using either edge $(s_2,t_1)$ or edge $(s_1,t_2)$ (depending on the realization of the edge costs). The next two iterations will therefore be degenerate. In one of these iterations $(s_2,t_1)$ is added to the spanning tree, but edge $(t_1,t_2)$ is saturated, prevents any flow being sent along the cycle, and is therefore removed from the spanning tree. In the other iteration $(s_1,t_2)$ is added to the spanning tree and $(s_2,s_1)$ is removed. After these two iterations the edges in $E_{UW}$ become eligible for augmenting flow in the reverse direction and the next $m$ iterations augment flow along the cycles $(t,q_{2M},\ldots,q_1,s)\cup P_{m+1}$, $\ldots$ , $(t,q_{2M},\ldots,q_1,s)\cup P_{2m}$.

Analogously to the above, every time an edge $(s_{i+1},s_{i})$ gets saturated, two degenerate iterations take place in which edges $(s_i,t_{i+1})$ and $(s_{i+1},t_i)$ are added to the spanning tree. This allows flow to be sent through $G_i$ in reverse direction, that is, from $t_i$ to $s_i$. Similarly, every time an edge $(t_{i+1},s_i)$ (that is, the reverse edge of an original edge $(s_i,t_{i+1})$) in the residual network gets saturated, two degenerate iterations take place in which edges $(t_i,t_{i+1})$ and $(s_{i+1},s_i)$ are added to the spanning tree. 

After $F$ iterations, there are no paths with positive residual capacity from $s_k$ to $t_k$. At this point another two degenerate iterations take place. In one of them edge $(c_1,t)$ is added to the spanning tree, but no flow is sent since $(s,a_1)$ has zero residual capacity and therefore $(s,a_1)$ leaves the spanning tree. In the other iteration $(s,b_1)$ is added to the spanning tree and $(d_i,t)$ leaves. Now the edges in $E_{UW}$ can be added to the spanning tree again and flow is augmented along them in reverse direction during the next $m$ iterations. In particular, in the next iteration flow is augmented along cycle $(t,q_{2M},\ldots,q_1,s)\cup P_{F+1}$.

Analogously to the above, every time an edge $(s,a_i)$ gets saturated, two degenerate iterations take place. In one of them $(c_i,t)$ enters the spanning tree and $(s,a_i)$ leaves. In the other $(s,b_i)$ enters the spanning tree and $(d_1,t)$ leaves. Also, every time an edge $(s,b_i)$ gets saturated, two degenerate iterations take place. In one of them $(d_{i+1},t)$ enters the spanning tree and $(s,b_i)$ leaves. In the other $(s,a_{i+1})$ enters the spanning tree and $(c_i,t)$ leaves.

Finally, after $2MF$ iterations none of the edges not in the spanning tree has both negative reduced cost and positive residual capacity and therefore the NS algorithm terminates. From the above discussion we can conclude that the NS algorithm on $G$ needs $2MF$ non-degenerate iterations plus several degenerate ones.

\begin{theorem} \label{NetworkSimplexG}
For flow network $G$ and initial spanning tree structure $(T,\tilde{L},\tilde{U})$, the NS algorithm needs $2MF$ non-degenerate iterations with probability $1$.
\begin{proof}
Follows immediately from the discussion above.
\end{proof}
\end{theorem}

Theorem~\ref{NetworkSimplexG} provides a lower bound for the number of iterations that the NS algorithm needs in the smoothed setting.

\begin{theorem}
For every $n$, every $m \in \{ n, \ldots, n^2 \}$, and every $\phi \leq 2^n$ there exists a flow network with $\Theta(n)$ nodes and $\Theta(m)$ edges, and an initial spanning tree structure for which the Network Simplex algorithm needs $\Omega(m\cdot \min\{n,\phi\}\cdot \phi)$ non-degenerate iterations with probability $1$.
\begin{proof}
Follows directly from Theorem~\ref{NetworkSimplexG} and the definition of $M$ and $F$.
\end{proof}
\end{theorem}

\section{Discussion}

In Section~\ref{LBNS} we showed a smoothed lower bound of $\Omega(m\cdot \min\{n,\phi\}\cdot \phi)$ for the number of iterations that the NS algorithm needs. This bound is the same as the smoothed lower bound that Brunsch et al.~\cite{BrunschSSP} showed for the SSP algorithm. For the SSP algorithm this lower bound is even tight in case $\phi=\Omega(n)$. Still, the NS algorithm is usually much faster in practice than the SSP algorithm. We believe that the reason for this difference is that the time needed per iteration is much less for the NS algorithm than for the SSP algorithm. In practical implementations, the entering edge is usually picked from a small subset (for example of size $\Theta(\sqrt{m})$) of the edges, which removes the necessity of scanning all edges for the edge which maximally violates its optimality conditions. Also, the spanning tree structure allows for fast updating of the flow and node potentials, in particular when the flow changes on only a small fraction of the edges. For the SSP algorithm an iteration consists of finding a shortest path, which takes $O(m+n\log(n))$ time. The experimental results of Kov\'acs~\cite{KovacsExperimental} seem to support this claim, since on all test instances the SSP algorithm is slower than the NS algorithm, but never more than a factor $m$. To allow a better comparison of the SSP algorithm and the NS algorithm in the smoothed setting, it would be useful to have a smoothed upper bound on the running time of the NS algorithm. Finding such an upper bound is our main open problem.

There is a gap between our smoothed lower bound of $\Omega(m\log(\phi))$ (Section~\ref{LBConstantPhi}) for the number of iterations that the MMCC algorithm requires and our smoothed upper bound of $O(mn^2\log(n)\log(\phi))$. Since our lower bound for the MMCC algorithm is weaker than the lower bound for the SSP algorithm, while the MMCC algorithm performs worse on practical instances than the SSP algorithm, we believe that our lower bound for the MMCC algorithm can be strengthened. Our stronger lower bound of $\Omega(mn)$ in case $\phi=\Omega(n^2)$ (Section~\ref{LBBigPhi}) is another indication that this is likely possible.

\end{document}